\documentclass{ws-p8-50x6-00}
\def\be{\begin{equation}}
\def\ee{\end{equation}}
\def\ba{$\begin{array}{c}}
\def\ea{\end{array}$}

\begin{document}

\title{hierarchical Majorana scales in the seesaw model}

\author{Guo-Hong Wu}

\address{Institute of Theoretical Science, University of Oregon,
Eugene, OR  97403, USA\\E-mail: wu@dirac.uoregon.edu}


\maketitle

\abstracts{
If the mixing angles in each of the seesaw sectors are all small,
and the neutrino masses are hierarchical, we study
the conditions for large neutrino mixing using triangular matrices.
In particular,  the implication of the
neutrino oscillation data for the mass hierarchy in the heavy Majorana
sector is examined, and 
  the heavy Majorana scales are shown to
depend sensitively on the solar mixing angle.}

\section{The Seesaw Mixing Matrix}

  The neutrino seesaw model\cite{ss},
$m_{\rm eff} = - m_D M^{-1} m_D^T $,
 involves both the Dirac and the Majorana 
mass matrices of which we have little information.
In grand unified theories, the Dirac neutrino mass matrix is often 
related to the quark mass matrix.  A similar analogy does not exist
for the heavy Majorana neutrino mass matrix. 
Here we will try to understand the implications of current neutrino
oscillation data\cite{atmos,solarnu,chooz} 
for the heavy Majorana sector\cite{kwm}.

   There are three components
 in the seesaw contributing
to the effective neutrino mixings:  left-handed (LH) rotations
in the charged lepton sector ($U_l$), LH rotations
in the Dirac neutrino sector ($U_d$), and 
right-handed (RH) mixings in the Dirac and heavy Majorana neutrino sectors.
The effective neutrino mixing matrix is defined as
 $\nu_i = U \nu_\alpha$, with $i=e,\mu, \tau$ and $\alpha=1,2,3$ for
the mass eigenstates.
We can write $U=U_l^\dag U_d U_{ss}$, where the seesaw mixing 
matrix\cite{ssen}
$U_{ss}$ is due to RH mixings in the Dirac and Majorana sectors: 
\be \label{eq1}
U_{ss}^\dag m^{\rm diag}_D V_R M^{-1}_{\rm diag} V_R^T m^{\rm diag}_D U_{ss}^*
= m^{\rm diag}_{\rm eff}  \; .
\ee
  Large neutrino mixing can be due to any of the three components 
or a certain combination of them\cite{ssen,phase,bando,afm,falcone,kwm}.
Here we will assume
  $U_l \simeq U_d \simeq 1$ so that we may identify $U$ with $U_{ss}$ and
investigate the restrictions on $V_R$ and the heavy Majorana scales
by present data. 
We will consider the limit of hierarchical neutrino masses in all sectors
and small RH mixing angles in  $V_R$.  
For numerical estimates, we neglect possible phases\cite{phase}
in the mass matrices and take the Dirac neutrino
masses to be the up type quark masses.

  Eq.~(\ref{eq1}) has a quadratic dependence on both $V_R$ and $U_{ss}$.
The analysis can be greatly simplified by reducing Eq.~(\ref{eq1}) to 
a linear equation. This is done by writing\cite{kwm} $m_{\rm eff}=N N^T$,
with $N=m^{\rm diag}_D V_R M^{-\frac{1}{2}}_{\rm diag}$, and dealing
with $N$ instead.  Denoting the mass eigenvalues of $M$ by
$M_i=R_i^{-2}$, the $N$ matrix is to a good approximation of lower
triangular form:
\bea \label{eq2}
N & \simeq & \left( \matrix{
       R_1m_1V_{11} & 0 & 0 \cr
       R_1m_2V_{21} & R_2m_2V_{22} & 0 \cr
       R_1m_3V_{31} & R_2m_3 V_{32} & R_3 m_3 V_{33} } \right)\; ,
\eea
where $m_i$ are the Dirac neutrino masses, and $V_{ij}$ are the 
elements of the RH mixing $V_R$.
$N$ has a linear dependence on $V_R$ and $R_i$ and will be our
center of focus. 
What do the neutrino data tell us about the $N$ matrix?

\section{General Solution}

   When we write $m_{\rm eff}=NN^T$, $N$ is defined up to an
arbitrary RH rotation $N \sim N O_R$.
This ambiguity can be removed by putting $N$ into lower
triangular form\cite{kwm,quarktri} through a properly chosen $O_R$.

  If $N=U_L {\rm diag}(n_1,n_2,n_3) U_R$, the effective neutrino
masses are given by $n_i^2$ and the mixing is $U=U_L$.
Thus all the information about effective neutrinos is contained in $N$.
Similarly, $N$ is uniquely determined in its lower triangular form 
by the masses and mixing angles of the effective neutrinos.
The properties of the heavy Majorana sector can then be read off
by identifying this triangular $N$ matrix with that of Eq.~(\ref{eq2}).

   The atmospheric neutrino data\cite{atmos}
 suggest that $\nu_\mu$-$\nu_\tau$ 
oscillation is consistent with maximal mixing.
Denoting the solar $\nu_e$-$\nu_\mu$ 
mixing angle by $\theta$ and $\nu_e$-$\nu_\tau$
mixing angle by $\epsilon$, the mixing matrix can be written
as $U=R_{23}(\pi/4) R_{13}(\epsilon) R_{12} (\theta)$.
The angle $\epsilon$ is small\cite{chooz}.

   Depending on the size of $\epsilon$, $N$  can have three
different patterns, with the largest elements
located in the (a) (2,2), (3,2), (b) $(i,j)$, $i,j=1,2$, (c) (2,1), (3,1)
positions, respectively.  However, only (a) does not involve fine-tuning
in the $N$ matrix.  Type (a) can be obtained in the limit
$n_1/n_2 \ll \theta$ and $\epsilon\ll m_2^{\rm eff}/m_3^{\rm eff}$. 
The $N$ matrix for (a) has a simple lower triangular form\cite{kwm}:
\be  \label{eq3}
N\simeq \left(\matrix{s_\theta n_2 & 0 &0\cr \frac{1}{\sqrt{2}}
 c_\theta n_2 &
\frac{1}{\sqrt{2}} n_3 & 0\cr
-\frac{1}{\sqrt{2}} c_\theta n_2 & \frac{1}{\sqrt{2}} n_3 &
\sqrt{2}\frac{n_1}{s_\theta}}\right)\; ,
\ee
where $c_\theta = \cos \theta$, and $s_\theta = \sin \theta$.

Comparison between  Eq.~(\ref{eq2}) and Eq.~(\ref{eq3})
gives the RH mixing angles:
\be
\frac{V_{32}}{V_{22}}=\frac{m_c}{m_t}, \;\;\;\;\;\;\;\;
\frac{V_{21}}{V_{11}}= \frac{m_u}{\sqrt{2}\tan \theta m_c},
\;\;\;\;\;\;\;\;
\frac{V_{31}}{V_{11}} =\frac{- m_u}{\sqrt{2}\tan \theta m_t}.
\ee 
All three RH angles are thus much smaller than one,
consistent with our assumption from the start.

 Identifying the diagonal elements of Eqs.~(\ref{eq2}) and (\ref{eq3}) 
yields three additional relations:
$m_3^{\rm eff} = 2m^2_c/M_2$,
$m_2^{\rm eff} = m^2_u/s^2_\theta M_1$, and
$m_1^{\rm eff} = s^2_\theta m^2_t/2 M_3$.
Note that the heaviest light neutrino is quadratically dependent on
$m_c$ rather than $m_t$.
For hierarchical effective neutrino masses, 
$m_3^{\rm eff} = \sqrt{ \Delta m^2_{\rm atm} }$ and
$m_2^{\rm eff} = \sqrt{ \Delta m^2_{\rm solar} }$.
The three heavy Majorana scales can then be obtained by reversing
the above three relations:
\be  \label{eqM2}
M_2  \simeq  \frac{2m_c^2}{\sqrt{\Delta m_{\rm atm}^2}}
\simeq 6\times 10^9 {\rm GeV} 
\ee 
and 
\be \label{eqM13}
\frac{M_1}{M_2}= \frac{m_u^2 m_3^{\rm eff}}{2 m_c^2 s_\theta^2
m_2^{\rm eff}}\; ,\;\;\;\;\;\;\;\;\;
\frac{M_2}{M_3}= \frac{4 m_c^2 m_1^{\rm eff}}{m_t^2 s_\theta^2
m_3^{\rm eff}}\; ,
\ee
where we have used $m_c(M_2) \approx 0.4$ GeV, and 
$\Delta m^2_{\rm atm} \approx 3 \times 10^{-3}$ eV$^2$ \cite{atmos}.
Note that different from $M_2$,
$M_1$ and $M_3$ depend on the solar mixing angle.

\section{Hierarchical Majorana Scales}

 We now examine the hierarchy
among the heavy Majorana scales for each of the solar solutions.
It is clear from Eq.~(\ref{eqM13}) that whereas 
$M_1$ may not be too far below the scale $M_2$, $M_3$ is always
far above $M_2$ independent of the particular solar solution.

\subsection{Small angle MSW}
Taking\cite{solarnu}
$\sin^2 2\theta_{\rm solar}\approx 5\times 10^{-3}$ and
$\Delta m^2_{\rm solar} \approx 5\times 10^{-6}\; {\rm eV}^2$,
we get
\be
M_1\approx 7\times 10^8 \; {\rm GeV} \;\;\;\;\;\;\;\;\;\;\;\;\;\;\;
M_3\approx r\cdot 4\times 10^{12} \; {\rm GeV}
\ee
where  $r=m^{\rm eff}_2/m^{\rm eff}_1 > 1$. 

\subsection{Large angle MSW}
Using\cite{solarnu} $\sin^2 2\theta_{\rm solar}\approx 0.8$ and
$\Delta m^2_{\rm solar} \approx 3\times 10^{-5}\; {\rm eV}^2$,
we have
\be
M_1\approx 1\times 10^6 \; {\rm GeV} \;\;\;\;\;\;\;\;\;\;\;\;\;\;\;
M_3\approx r\cdot 4\times 10^{14} \; {\rm GeV} \; .
\ee

\subsection{Vacuum oscillation}
Taking\cite{solarnu} $\theta_{\rm solar}\approx 45^\circ$ 
and $\Delta m^2_{\rm solar} \approx 7\times 10^{-11}\; {\rm eV}^2$,
we find
\be
M_1\approx 5\times 10^8 \; {\rm GeV} \;\;\;\;\;\;\;\;\;\;\;\;\;\;\;
M_3\approx r\cdot 4\times 10^{17} \; {\rm GeV} \; .
\ee
The scale $M_3$ being way above $M_{\rm GUT}$
makes the vacuum oscillation solution unlikely to be viable.

  For all three solutions, we find a hierarchy among the heavy Majorana
scales $M_1 \ll M_2 \ll M_3$, and the separation between scales depends
sensitively on the specific solution to the solar neutrino problem.
The same conclusion was reached in a different way\cite{falcone}.

In summary, though it is possible to get large neutrino mixing
from hierarchical masses and small mixing angles in all of the
seesaw components, a strong hierarchy of heavy Majorana scales seems
to be necessary.

\section*{Acknowledgments}
 I would like to thank T.K. Kuo and S. Mansour
for enjoyable collaboration.

\end{document}